\documentclass[aps,nofootinbib,11pt,preprintnumbers,prd]{revtex4-1}
\usepackage{amssymb,amsmath}
\usepackage{graphics}
\usepackage{graphicx}
\usepackage{color}

\usepackage{todonotes}

\newcommand{\bea}{\begin{eqnarray*}}
\newcommand{\eea}{\end{eqnarray*}}

\newcommand{\beq}{\begin{equation}}
\newcommand{\eeq}{\end{equation}}
\newcommand{\eq}[1]{Eq.~(\ref{#1})}

\begin{document}

\title{
Two-dimensional chiral crystals in the NJL model
}

\author{Stefano~Carignano}
\author{Michael~Buballa}

\affiliation{
Institut f\"ur Kernphysik (Theoriezentrum), 
Technische Universit\"at Darmstadt, Germany
}

\date{\today}

\begin{abstract}
We investigate the phase structure of the Nambu--Jona-Lasinio model 
at zero temperature, allowing for a two-dimensional
spatial dependence of the chiral condensate.
Applying the mean-field approximation,
we consider various periodic structures with rectangular and hexagonal 
geometries, and minimize the corresponding free energy. 
We find that these two-dimensional chiral crystals are favored over 
homogeneous phases in a certain window in the region where the phase 
transition would take place when the analysis was restricted to 
homogeneous condensates.
It turns out, however, that in this regime they are disfavored against a 
phase with a one-dimensional modulation of the chiral condensate.
On the other hand, we find that square and hexagonal lattices eventually 
get favored at higher chemical potentials.
Although stretching the limits of the model to some extent, this would 
support predictions from quarkyonic-matter studies.
\end{abstract}

\maketitle

\section{Introduction}

Since more than three decades, the phase diagram of quantum chromodynamics
is object of intensive theoretical and experimental 
research~\cite{Halasz:1998qr,Stephanov,BraunMunzinger:2008tz,Fukushima:2010bq}. 
In particular the conjecture that at low temperatures there could be a 
first-order chiral phase transition which ends at a critical point
has received a lot of interest, since this endpoint is potentially
detectable in heavy-ion experiments.
However, in most studies which support this scenario,
it is tacitly assumed that the order parameters
of the various phases are uniform in space.
On the other hand, it is not a new idea that there could be 
spatially modulated states in strongly interacting matter 
(see Ref.~\cite{Broniowski:2011} for a recent review).
Well-known examples are 
the proposal of an inhomogeneous ground state in nuclear 
matter~\cite{Overhauser:1962},
the possibility of spatial modulations in the context of pion 
condensation~\cite{Dautry:1979,Broniowski:1990},
Skyrme crystals~\cite{Goldhaber:1987}, 
and crystalline phases in (color-) 
superconductors~\cite{Fulde:1964,LO64,Matsuda:2007,Alford:2000ze,Bowers:2002,Casalbuoni:2004,Rajagopal:2006,NB:2009}.

For the QCD phase diagram, 
it has been argued some time ago that, at least in the limit of a large number of colors 
($N_c$), the favored ground state of a dense
Fermi sea of quarks should be characterized by a spatial modulation 
of the chiral condensate~\cite{Deryagin:1992,Shuster:1999}.
More recent studies on quarkyonic matter
seem to support this hypothesis~\cite{Kojo:2009,Kojo:2010,Kojo:2011}.

For the physical case of three colors, Nambu--Jona-Lasinio-- (NJL--) type 
model studies have revealed 
at intermediate chemical potentials and low temperatures 
the presence of an inhomogeneous phase
where the chiral condensate assumes a spatially modulated 
form~\cite{Sadzikowski:2000,NT:2004,Nickel:2009}.
Most of the existing studies on inhomogeneous phases have restricted their 
analysis to simplified shapes of the chiral order parameter.
The most popular example is the so-called ``chiral density wave'' (CDW), 
which basically amounts to a single plane 
wave~\cite{Sadzikowski:2000,NT:2004,Frolov:2010}.
This kind of ansatz is
analogous to the so-called Fulde-Ferrel solutions
proposed in (color-) superconductivity~\cite{Fulde:1964}.

The study of a generic shape for the spatially dependent chiral condensate
is a highly non-trivial task.
It has recently been observed, however, that in NJL-type models
the evaluation of the energy spectrum simplifies considerably 
when the condensate is allowed to vary only in one spatial dimension,
while remaining constant along the two transverse 
directions \cite{Nickel:2009}.
In this particular case, the problem can be reduced to the 
1+1-dimensional chiral Gross-Neveu model, where analytical 
expressions for the eigenvalue spectra
are known \cite{Schnetz:2004,Schnetz:2005,Thies:2006,Basar:2009fg}.
This formal resemblance allows to perform an analysis of 
the phase diagram including these
inhomogeneous phases without having to calculate
numerically the energy spectrum of the model.

Within this framework it has been found that the inhomogeneous phase
covers the region where a first-order chiral phase transition would occur 
when limiting the analysis to homogeneous phases. As a consequence 
the chiral critical point disappears from the phase diagram,
leaving only a Lifshitz point where three second-order lines 
meet~\cite{Nickel:2009,Nickel:2009prl}.
The inclusion of vector interactions further enhances this effect
and enlarges the size of the inhomogeneous phase \cite{CNB:2010}.

The limitation to one-dimensional structures is of course a strong one. 
Especially at lower temperatures,
higher dimensional modulations are expected to play an important role 
in (color-) superconductors~\cite{Matsuda:2007,Bowers:2002}, 
whose dynamics bears strong formal resemblance 
with the one described in the NJL model.
Moreover, recent quarkyonic-matter studies suggest that
in the high chemical potential region,
as the density of the system grows,
the quark Fermi sea tends to break chiral symmetry by forming
increasingly complex crystalline (or quasi-crystalline) structures, 
which can be described as superpositions of several
``quarkyonic chiral spirals'' \cite{Kojo:2010,Kojo:2011}.

Aside from these considerations,
modulations in more than one spatial dimension are also of interest
since they are
unaffected by the instabilities with respect to fluctuations which prevent
the formation of a true 1D crystalline structure at finite temperature
\cite{Baym:1982}.

The main purpose of this paper is therefore to investigate the properties of modulations
of the chiral condensate occurring in more than one spatial dimension. While a complete analysis would in principle 
have to allow for modulations in three spatial dimensions, for computational reasons we will
limit ourselves in the present work to two-dimensional crystalline shapes. The generalization
to three-dimensional structures is straightforward once the formalism
and the numerical procedure are set up.

\section{Inhomogeneous phases in the NJL model}
\label{sec:model}
Our starting point is the two-flavor NJL Lagrangian~\cite{NJL2},
\begin{equation}
\mathcal{L}_{NJL} = \bar\psi (i\gamma^\mu\partial_\mu -m)\psi + G \left( ({\bar\psi\psi})^2 + (\bar\psi i \gamma^5 \tau_a \psi)^2\right)\,,
\end{equation}
where $\psi$ is a quark field with 
two flavor and three color degrees of freedom and
bare mass $m$,
$\tau_a$ denotes the three Pauli matrices in isospin space,
and $G$ is a coupling constant.

We perform the mean-field approximation by expanding the interaction
around the scalar and pseudoscalar condensates
\begin{equation}
\langle\bar{\psi}\psi\rangle=S(\vec x)  \,,\qquad 
\langle\bar{\psi}i\gamma^5\tau^a\psi\rangle=P(\vec x)\,\delta_{a3} \,,
\end{equation}
which we allow to be space dependent.
For later convenience, we also introduce the complex ``mass'' function,
\begin{equation}
M(\vec x) = m - 2G (S(\vec x) + i  P(\vec x)) \,.
\end{equation}
The mean-field Lagrangian can then be written as
\begin{equation}
\mathcal{L_{MF}} = 
\bar\psi \gamma^0 (i\partial_0 - \mathcal{H})\psi - G (S^2 + P^2) \,,
\label{eq:LMF}
\end{equation}
with the effective Hamiltonian operator
\begin{equation}
\mathcal{H} =  \gamma^0 \left[ i\vec\gamma\cdot\vec\partial + m - 2G(S + i \gamma^5 \tau_3 P) \right]\,.
\end{equation}
The mean-field thermodynamic potential per volume $\Omega$ associated with
these (so far generic) spatially modulated condensates at temperature $T$ and 
quark chemical potential $\mu$ contains a functional trace over the
logarithm of the inverse quark propagator \cite{Kap:89}.
For its evaluation we employ imaginary time formalism and switch to 
momentum space.
Assuming static (i.e., time-independent) condensates, we can perform the 
sum over Matsubara frequencies explicitly and obtain
(up to a constant)
\begin{equation}
\label{eq:Omega2}
\Omega(T,\mu;M(\vec x))
=
\Omega_\mathit{kin}(T,\mu;M(\vec x))
+
\Omega_\mathit{cond}(M(\vec x))
\,,
\end{equation}
with
\begin{equation}
\label{eq:Omegacond}
\Omega_\mathit{cond}(M(\vec x))
=
\frac{1}{V} \int_V\, d^3x \,\frac{\vert M(\vec x)-m\vert^2}{4G}
\,,
\end{equation}
where $V$ is the volume of the system, and
\begin{equation}
\label{eq:Omegakin}
\Omega_\mathit{kin}(T,\mu;M(\vec x))
=
-T
\sum_{E}
\log\left(2\cosh\left(\frac{E-\mu}{2T}\right)\right)
\,,
\end{equation}
where the sum runs over all eigenvalues $E$ of $\mathcal{H}$ in color, 
flavor, Dirac and momentum space.

In presence of an inhomogeneous condensate,
the diagonalization of 
$\mathcal{H}$ is a highly non-trivial task,
since quarks may exchange momenta by scattering off the 
condensate
and, consequently, the resulting mean-field quark propagator 
is not diagonal in momentum space.
In the following we assume
a periodic shape of the chiral condensate forming a well-defined lattice 
structure.
This implies that we can expand the spatially varying 
order parameter in a Fourier series,
\begin{equation}
M(\vec x) = \sum_{\vec{q}_k} M_{\vec{q}_k} e^{i \vec{q}_k\cdot\vec x} \,,
\label{eq:Mq}
\end{equation}
with discrete momenta $\vec{q}_k$ forming a reciprocal lattice (RL). 
A generic element of $\mathcal{H}$ in momentum space then takes the form
\begin{equation}
\mathcal{H}_{\vec{p}_m,\vec{p}_n} =
 \left( 
\begin{array}{cc}
 -\vec\sigma\cdot\vec{p}_m\,\delta_{\vec{p}_m,\vec{p}_n} &  
 \sum_{\vec{q}_k} M_{\vec{q}_k} \delta_{\vec{p}_m,\vec{p}_n+\vec{q}_k} 
 \\
 \sum_{\vec{q}_k} M^*_{\vec{q}_k} \delta_{\vec{p}_m,\vec{p}_n-\vec{q}_k} &  
 \vec\sigma\cdot\vec{p}_m\,\delta_{\vec{p}_m,\vec{p}_n} 
\end{array} 
\right) \,, 
\label{eq:Hmn}
\end{equation}
where $\sum_{\vec{q}_k}$ runs over the momenta of the RL, making obvious
the non-diagonal momentum structure of the matrix.\footnote{The matrix is 
also non-diagonal in Dirac space, as indicated in 
\eq{eq:Hmn}. Here the chiral representation was used, and $\vec\sigma$ 
corresponds to the Pauli matrices. On the other hand, $\mathcal{H}$ is 
diagonal in isospin and color.}
In turn, momenta which do not differ by an element of the RL are not
coupled, so that $\mathcal{H}$ can be decomposed into a block diagonal
form, where each block $\mathcal{H}(k)$ can be labelled by an element of the first
Brillouin zone (BZ).
This implementation of Bloch's theorem
allows to decompose the eigenvalue sum in Eq.~(\ref{eq:Omegakin}) into 
a momentum integration over the BZ times a sum over the discrete eigenvalues 
of each block \cite{NB:2009}.

\section{Two-dimensional modulations}
\label{sec:twodim}
For general periodic structures, 
although the numerical diagonalization procedure is in principle 
straightforward,
its practical implementation turns out to be computationally demanding.
In order to simplify the problem, we therefore limit the generality 
of our ansatz Eq.~(\ref{eq:Mq}) to lower-dimensional modulations.
In this case, the 
momentum integration required for the evaluation of the thermodynamic 
potential may be split into the parts $\vec{p}_\parallel$
along the direction of the modulation and $\vec{p}_\perp$ perpendicular 
to it. One thus obtains, for a $d$-dimensional modulation, 
\begin{equation}
\Omega_\mathit{kin}
= -\!\int \frac{d^{3-d}p_\perp}{(2\pi)^{3-d}}\int\limits_{BZ} \frac{d^{d}k}{(2\pi)^{d}} \sum_{E} T \log \left[ 2\cosh \left(\frac{E(\vec{p}_\perp,\vec{k}) - \mu}{2T}\right) \right],
\label{eq:omega2d}
\end{equation}
where $\vec{k}$ labels the BZ momenta and
$E(\vec{p}_\perp,\vec{k})$ are the eigenvalues of $\mathcal{H}(\vec{k})$ for a 
given $\vec{p}_\perp$.

It has been observed that the eigenvalues in 3+1 dimensions
can be evaluated by 
diagonalizing a dimensionally reduced $\mathcal{H}$ 
(evaluated at $p_\perp = 0$)
and subsequently boosting the resulting spectrum~\cite{Nickel:2009}. 
This dramatically simplifies the calculations. 
In particular for the case of one-dimensional modulations it
allows to reuse a well-established set of analytical results without having to 
resort to a numerical diagonalization of the model Hamiltonian. 
The favored mass functions are then given by Jacobi elliptic 
functions,  which smoothly interpolate between solitonic shapes close to the 
homogeneous chirally broken phase and sinusoidal shapes close to the restored 
phase.

Since the one-dimensional problem has already been treated extensively
in previous works, the main focus of this work is on two-dimensional 
structures.
A recent Ginzburg-Landau (GL) analysis has shown that 
close to the Lifshitz point one-dimensional modulations 
are energetically favored over higher dimensional ones
in NJL-type models~\cite{Abuki:2011}.
We therefore focus on what happens at zero temperature, 
where GL arguments are unable to provide reliable results.
For simplicity, we restrict our calculations to the chiral limit, $m=0$.

Without loss of generality, we assume the chiral condensate to vary in the
$xy$-plane and to be constant in the $z$-direction. 
Unlike for the one-dimensional case, 
in two spatial dimensions different crystalline shapes may be realized.
Therefore we consider different lattice geometries
and assume the mass functions to have simple symmetric shapes consistent
with these structures.

\begin{figure}
\begin{center}
\includegraphics[height=.4\textwidth,angle=270]{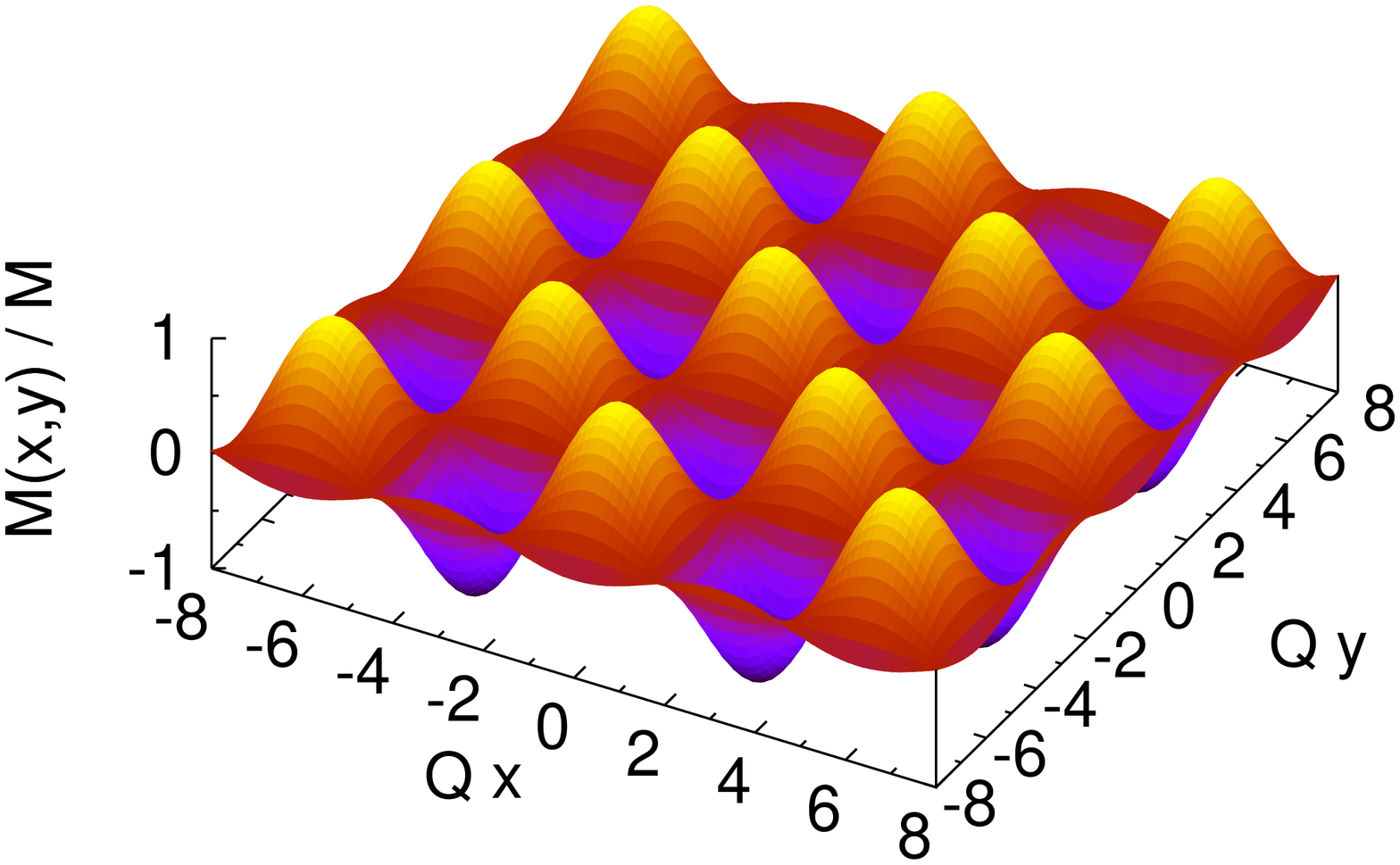}
\includegraphics[height=.4\textwidth,angle=270]{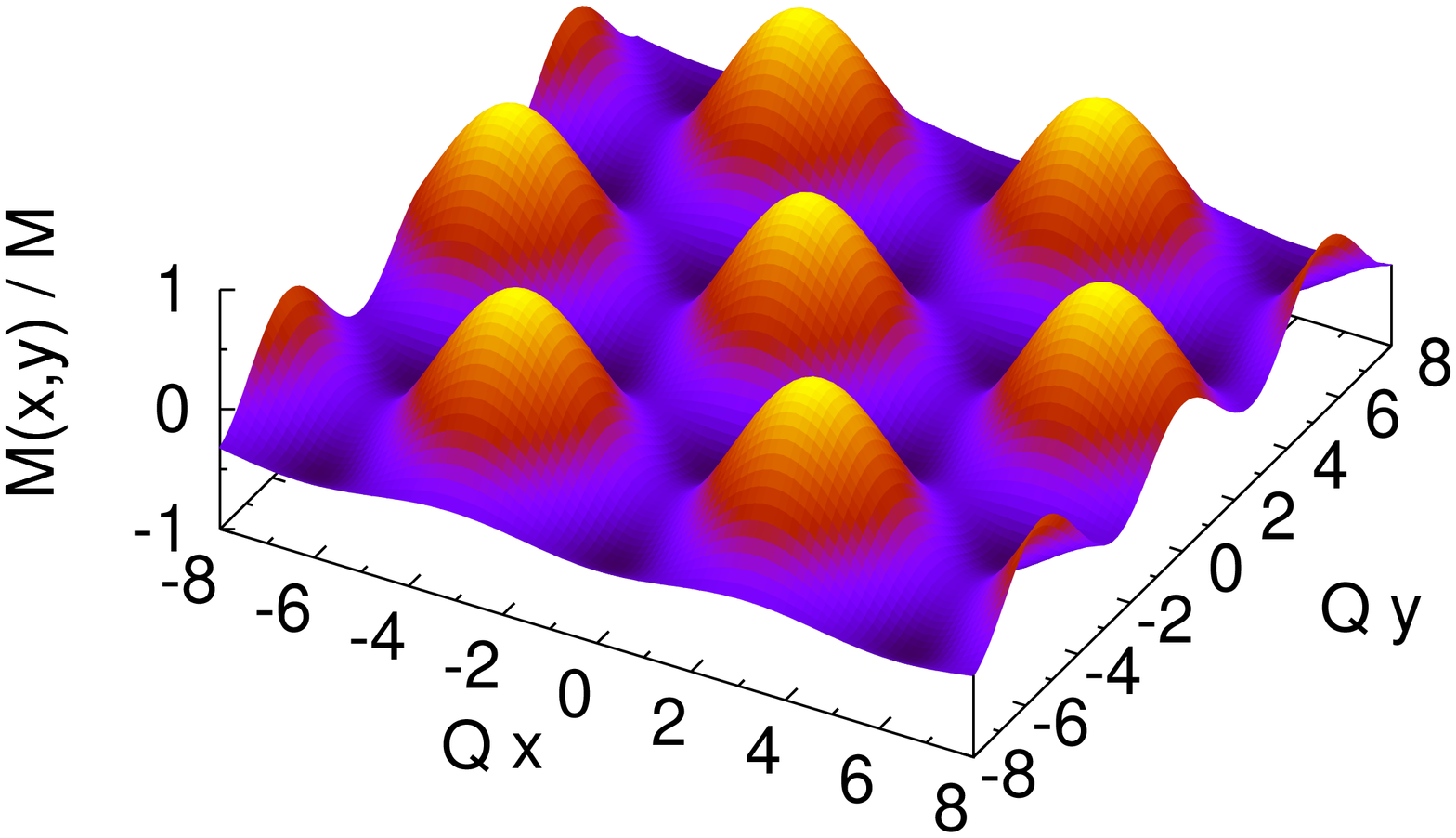}
\end{center}
\caption{Mass functions $M(x,y)$
with two-dimensional modulations in coordinate space.
Left: ``egg-carton'' modulation on a square lattice, \eq{eq:Mcos2d}. 
Right: hexagonal modulation, \eq{eq:Mhex}.}
\label{fig:2dmods}
\end{figure}

The first case is a square lattice with a unit cell spanned by 
two perpendicular vectors of length $a$ in $x$ and $y$ direction.
The corresponding elements of the RL are then given by 
$\vec{q}_{m,n} = Q(m\vec{e}_x + n\vec{e}_y)$ with $Q = 2\pi/a$ and
integers $m$ and $n$.
While the general mass function consistent with this lattice structure would 
be given by \eq{eq:Mq} with arbitrary Fourier coefficients $M_{m,n}$,
we restrict ourselves to a simple ansatz for a real symmetric mass function 
with a small number of nonvanishing Fourier coefficients. 
Specifically we choose
$M_{1,1} = M_{1,-1} = M_{-1,1} = M_{-1,-1}= M/4$ and $M_{m,n} = 0$ in all
other cases.\footnote{An alternative choice would be
$M_{1,0} = M_{-1,0} = M_{0,1} = M_{0,-1}= M/4$ and $M_{m,n} = 0$ in all
other cases, which yields $M(x,y) = \frac{M}{2} (\cos(Q x)+\cos(Q y))$. 
However, this is equivalent to \eq{eq:Mcos2d} in a frame rotated by 
$\pi/4$ and $Q$ replaced by $Q/\sqrt{2}$.} 
This yields
\begin{equation}
  \label{eq:Mcos2d}
 M(x,y) = M \cos(Q x)\cos(Q y) \,,
\end{equation}
which has an egg-carton-like shape (see Fig.~\ref{fig:2dmods}, left)
and is symmetric under discrete rotations by $\pi/2$.

The second case we consider is a mass function with hexagonal symmetry.
Here we start from a unit cell spanned by two vectors of length $a$, enclosing
an angle of $\pi/3$. Choosing the first one to be aligned with the $x$-axis,
the elements of the RL are given by 
$\vec{q}_{m,n}= Q (m \vec{e}_x  + \frac{2n-m}{\sqrt{3}}\vec{e}_y)$,
with integers $m$ and $n$, and $Q = 2\pi/a$ as before.
For the mass function we choose 
$M_{m,n} = M/6$ on the corners of a regular hexagon,
$(m,n) \in \{(1,0), (-1,0), (0,1), (0,-1), (1,1), (-1,-1)\}$,
and $M_{m,n} = 0$ in all other cases. 
This yields 
\beq
\label{eq:Mhex}
M(x,y) = 
 \frac{M}{3} \left[ 
2 \cos\left(Q x\right) \cos\left( \frac{1}{\sqrt{3}} Qy\right) 
+
\cos(\frac{2}{\sqrt{3}} Q y) 
\right] \,,
\eeq
which is symmetric under discrete rotations by $\pi/3$ 
(see Fig.~\ref{fig:2dmods}, right).
Note that the normalization of the amplitude was chosen to 
match the homogeneous case $M(x,y) = M$ 
when $Q$ goes to zero.

After inserting these shapes into \eq{eq:Hmn}, 
the Hamiltonian is diagonalized numerically in Dirac and momentum space, 
and the thermodynamic potential 
is minimized with respect the variational parameters $M$ and $Q$,
characterizing amplitude and period of the modulations.
For the numerical calculations, we have to specify a procedure to 
regularize the integrals and eigenvalue sum in \eq{eq:omega2d}.
Following Refs.~\cite{Nickel:2009,CNB:2010,CB:2011}, we use 
a Pauli-Villars scheme with three regulators. 
For the results in this section, we fit our model parameters 
(the regulator $\Lambda$ and the coupling constant $G$)
to reproduce the pion decay constant in the chiral limit, $f_\pi = 88$ MeV,
and a vacuum constituent quark mass of $M_{vac}=300$ MeV \cite{Nickel:2009}.

\begin{figure}
  \begin{center}
  \includegraphics[height=.4\textwidth,angle=270]{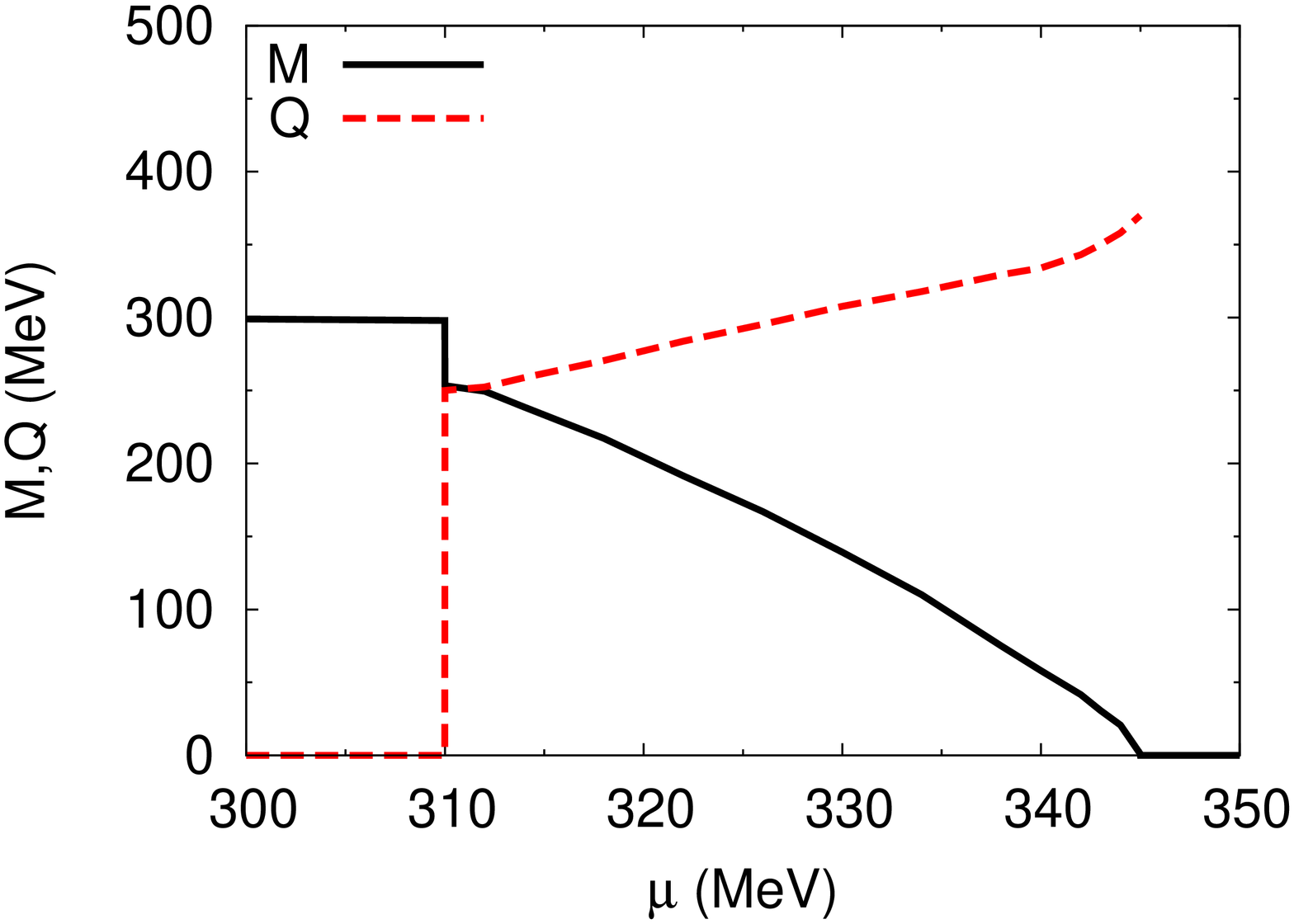}
  \includegraphics[height=.4\textwidth,angle=270]{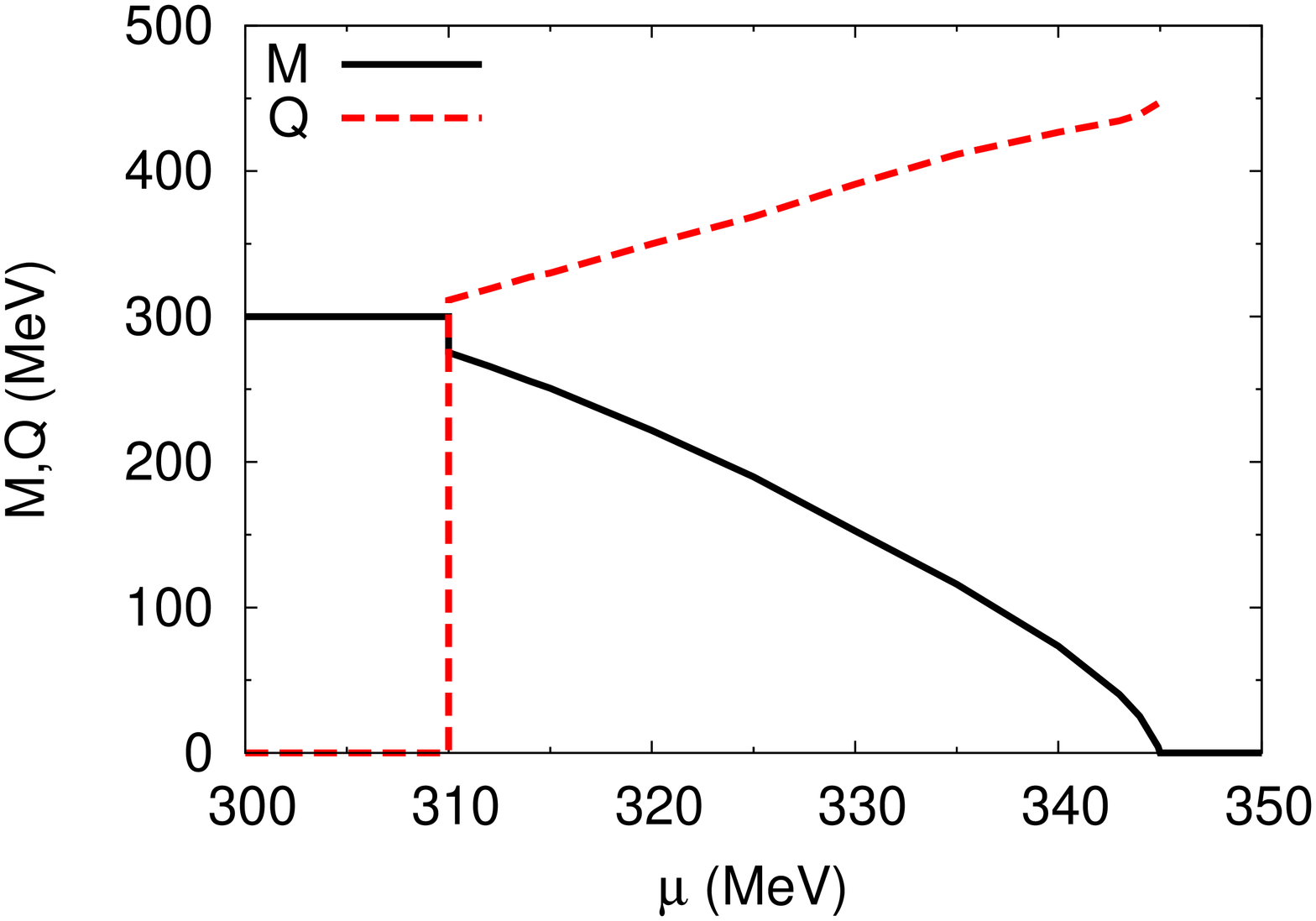}
  \end{center}
  \caption{Amplitude $M$ and wave number $Q$ at $T=0$ as functions of
  the chemical potential $\mu$ after minimization of the thermodynamic 
  potential for given shapes of the mass function.
  Left: ``egg-carton'' modulation on a square lattice, \eq{eq:Mcos2d}. 
  Right: hexagonal ansatz, \eq{eq:Mhex}. 
  }
  \label{fig:tpmin2d}
\end{figure}

The results of the numerical minimization for the square and hexagonal shapes,
Eqs.~(\ref{eq:Mcos2d}) and (\ref{eq:Mhex}) respectively, are presented in 
Fig.~\ref{fig:tpmin2d}.
For both modulations, we find a sharp onset of the crystalline phase
around $\mu \approx 310$ MeV and a smooth 
approach to the restored phase, which is reached at $\mu \approx 345$ MeV
via a second-order phase transition as the amplitude 
of the chiral condensate melts to zero. 
For the transition to the homogeneous chirally broken phase, 
the situation is therefore different from
the case of one-dimensional solitonic
solutions \cite{Nickel:2009} and might be due to our limited ansatz 
with a finite number of Fourier components. 
We also note that the results for the amplitude $M$  
are comparable for both shapes in the inhomogeneous window.

The results presented in Fig.~\ref{fig:tpmin2d}
have been obtained by enforcing in each case a fixed shape of the chiral 
modulation.
Under this restriction we found a window where the different two-dimensional 
solutions are energetically favored over the homogeneous solutions.
This is basically the same region where one-dimensional modulations
are also found to be favored over homogeneous solutions.
In fact, as shown in Ref.~\cite{Nickel:2009prl} using GL arguments,
the critical chemical potential for the transition
from the inhomogeneous to the chirally restored phase is independent of
the shape of the spatial modulation, as long as the phase transition is 
second order.

\begin{figure}
\begin{center}
\includegraphics[width=.48\textwidth]{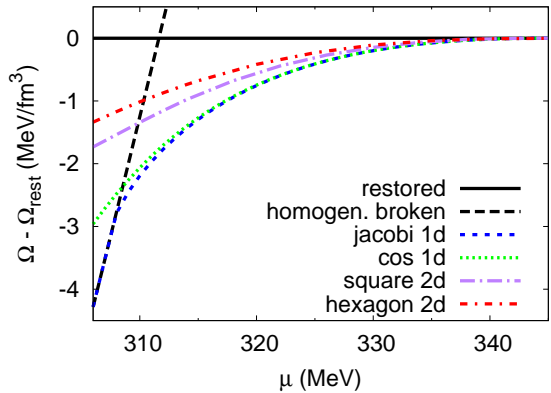}
\end{center}
\caption{
Thermodynamic potential relative to the restored phase
for different modulations of the chiral condensate 
at $T=0$. 
The homogeneous broken and restored solutions are disfavored compared to 
the crystalline phases in a window between $\mu \approx 308$ and
$\mu \approx 345$ MeV. 
The lowest free energy is found for the one-dimensional Jacobi elliptic 
function. 
In particular at the onset of the inhomogeneous phase it is
favored over a one-dimensional cosine, while with increasing $\mu$
the two shapes quickly become almost degenerate.
The two-dimensional square ansatz, \eq{eq:Mcos2d}, 
is always disfavored against the one-dimensional modulations, 
and the hexagonal crystal, \eq{eq:Mhex},
leads to an even smaller gain in free energy.} 
\label{fig:cfrtp}
\end{figure}

The next obvious step is to compare the free energies
of these solutions with each other, in order to find out
which of them corresponds to the most stable solution.
The results of our comparison are shown in Fig.~\ref{fig:cfrtp}.
One can clearly see that the one-dimensional Jacobi elliptic functions
lead to the biggest gain in free energy compared to all the other cases 
considered.
In particular, the two-dimensional structures turn out to be 
energetically disfavored with respect to one-dimensional real modulations 
throughout the whole inhomogeneous window.

\begin{figure}
\begin{center}
\includegraphics[width=.65\textwidth]{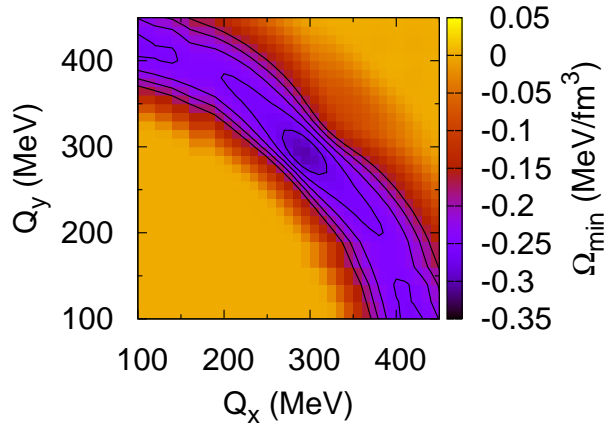}
\end{center}
\caption{Value of the thermodynamic potential at $T=0$ and $\mu=325$~MeV for 
a rectangular lattice, \eq{eq:Mcos2drec}, as a function of $Q_x$ and $Q_y$. 
At each point, $\Omega$ was minimized with respect to $M$.}
\label{fig:omegaPxPy}
\end{figure}

In this context it is instructive to introduce a rectangular structure,
which interpolates continuously between a square lattice 
and a one-dimensional periodic modulation.
Specifically, we can generalize the ``egg-carton'' ansatz \eq{eq:Mcos2d} 
to
\begin{equation}
  \label{eq:Mcos2drec}
 M(x,y) = M \cos(Q_x x)\cos(Q_y y) \,,
\end{equation}
which reduces to a single cosine varying in one spatial dimension when one 
of the two wave numbers goes to zero. 

Starting from this ansatz, we minimize the thermodynamic potential
with respect to the amplitude $M$ for fixed wave numbers $Q_x$ and $Q_y$, 
and then study the result as a function of $Q_x$ and $Q_y$. 
Since we know already that the square-lattice solution along the line
$Q_x = Q_y$ is disfavored against the one-dimensional cosine, 
we are mainly interested in the question whether the former corresponds 
to a local minimum or to a saddle point in the $Q_x-Q_y$ plane.

Our result for $\mu = 325$~MeV is presented in Fig.~\ref{fig:omegaPxPy},
showing that the solution at $Q_x = Q_y$ is a local minimum.
One can also see that in the vicinity of the minimum the potential
remains rather flat along the direction perpendicular to the line 
$Q_x = Q_y$. 
Going along this valley, we find two saddle points at 
$(Q_x,Q_y) \approx (175\,\mathrm{MeV},400\,\mathrm{MeV})$ and 
$(400\,\mathrm{MeV},175\,\mathrm{MeV})$. 
Unfortunately, the computing time rises strongly with decreasing wave
numbers, so that we could not continue this analysis to values
of $Q_x$ or $Q_y$ lower than 100~MeV. 
However, it is not hard to imagine how beyond the
saddle points the valley approaches the absolute minima at vanishing
$Q_x$ or $Q_y$, corresponding to a one-dimensional cosine.

One may ask whether our observation that the two-dimensional crystalline 
structures are disfavored against the one-dimensional ones is caused by 
the restricted ansatz for the mass functions. 
Taking into account more Fourier modes would lead to additional
variational parameters and could thus lower the free energy.
It is however unlikely that this would change our results considerably.
As seen in Fig.~\ref{fig:cfrtp} the difference in free energy between 
the Jacobi elliptic function and the one-dimensional cosine is
negligible in a large chemical-potential range and always
much smaller than the difference to the two-dimensional solutions.
We therefore expect that the corrections to the considered two-dimensional
shapes are small as well.

In order to test this, we extend the simple ``egg carton'' ansatz 
(Eq.\ref{eq:Mcos2d}) into
\begin{equation}
   M(x,y) = \sum_{n=1}^3 M_n \cos(nQx)\cos(nQy) 
\end{equation}
The minimization of the thermodynamic potential with respect to 
the variational parameters ($M_1, M_2, M_3, Q$) leads,  
within numerical errors, to $M_2 = M_3 = 0$ throughout the whole 
inhomogeneous window. 
Although numerical errors are more significant 
than for the one-dimensional case, it is safe to say that
the inclusion of those higher harmonics considered above
does not lead to an appreciable gain in free energy.

\section{Higher chemical potentials}
\label{sec:cont}

In recent quarkyonic-matter studies it was found that 
increasing the chemical potential leads to two-dimensional
structures with growing geometrical complexity~\cite{Kojo:2011}.
While no definite scale was given, this could be a hint that the 
chemical potentials we have considered so far are too low for 
two-dimensional crystals to be favored.
This motivates us to extend our investigations to higher values of $\mu$.

In fact, while so far we have concentrated on the inhomogeneous ``island''
close to the would-be first-order phase boundary for homogeneous phases,
we have recently found that in the NJL model a second inhomogeneous
region (``continent'') appears at higher $\mu$ and seemingly  
persists to arbitrarily high chemical potentials~\cite{CB:2011}.
Of course, we have to keep in mind that the NJL model is a low-energy
effective model with a limited range of validity.  
In particular, since the continent appears in a region where the 
chemical potential is of the order of the regulator masses, 
we have to be cautious not to overinterpret the results. 
However, as thoroughly discussed in Ref.~\cite{CB:2011}, 
although the inhomogeneous continent is sensitive to regularization
effects, it is not obviously created by them.
Indeed, there is no a priori reason to exclude the possibility of an 
inhomogeneous chiral symmetry breaking phase at high chemical 
potentials. 
For instance,
inhomogeneous phases extending to arbitrarily high 
chemical potentials have been predicted for the Gross-Neveu model and its 
chiral counterpart~\cite{Schnetz:2004}, for quarkyonic matter~\cite{Kojo:2011}
and for QCD in the large-$N_c$ limit~\cite{Deryagin:1992}.
Here we do not want to enter this discussion, but simply take the
inhomogeneous continent as a ``model laboratory''
to study the competition of one- and two-dimensional chiral crystals
as a function of $\mu$. 

\begin{figure}
\begin{center}
\includegraphics[height=.48\textwidth,angle=270]{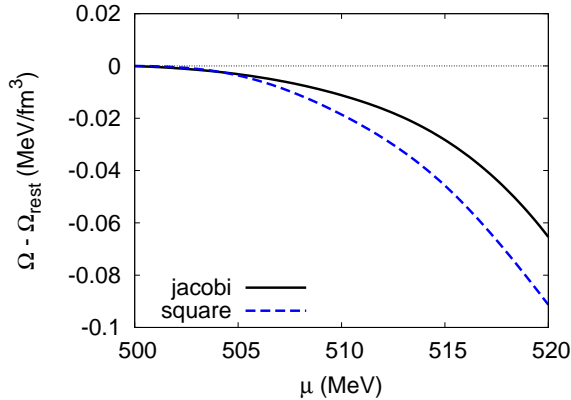}
\end{center}
\caption{Free energies
  associated with the one-dimensional solitonic solutions (black solid line)
  and the two-dimensional ``egg-carton'' (blue dashed line) at the onset of the second inhomogeneous phase,
  occurring around $\mu = 500$ MeV.
  Results are calculated for $T=0$ with a parameter set fitted 
  to give a vacuum constituent quark mass of $M_{vac} = 300$ MeV, and are normalized
  with respect to the free energy of the restored phase.
}
\label{fig:onset}
\end{figure}

\begin{figure}
\begin{center}
\includegraphics[height=.48\textwidth,angle=270]{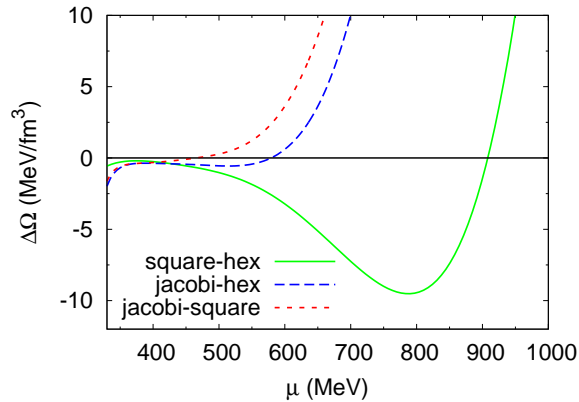}
\end{center}
\caption{Free-energy differences between inhomogeneous phases with
different modulations, evaluated at $T=0$ and with a parameter set 
fitted to give a vacuum constituent quark mass of $M_{vac} = 330$ MeV.
}
\label{fig:cfrtphiM}
\end{figure}

Figure \ref{fig:onset} shows the free energies associated with 
the one-dimensional solitonic solutions and the
two-dimensional ``egg carton'' close to the 
phase transition marking the onset of the inhomogeneous continent 
(the second-order nature of the transition from the restored phase is visible in the behaviour of 
  both free energies).
From this comparison, it is possible to see that in this
region, two-dimensional solutions become favored over one-dimensional ones.
Since with the current parameter set the continent is not connected
to the the inhomogeneous island, it is not clear at which point higher-dimensional
structures become favored. In order to achieve a better understanding of the 
problem, in the following we therefore employ a 
%
%
slightly modified parameter set with a vacuum 
constituent quark mass of $330$~MeV. This does not modify any qualitative
behaviour of the model but has the advantage that the 
inhomogeneous island merges with the continent, so that the comparison
of the inhomogeneous phases can be performed on a continuous interval, 
without being interrupted by the restored phase.
  
In Fig.~\ref{fig:cfrtphiM} the differences between the free energies
of three inhomogeneous phases with different modulations are displayed 
as functions of $\mu$.
As we have seen before, at low chemical potentials the two-dimensional 
crystals are disfavored against the one-dimensional Jacobi elliptic 
function. 
Above $\mu \approx 450$~MeV, however, the two-dimensional square lattice 
leads to a lower free energy. 
At $\mu \approx 600$~MeV, also the hexagon surpasses the 
one-dimensional modulation and finally becomes the most favored shape at 
$\mu \approx 900$~MeV. 
Thus, while being aware that the model cannot be trusted blindly in this
high density region, it is nevertheless remarkable that we recover the 
same sequence of crystalline phases as described in Ref.~\cite{Kojo:2011}.

\section{Discussion and outlook}
\label{sec:concl}

In this paper 
we presented the results of our numerical study of two-dimensional 
chiral crystalline structures in the NJL model at zero temperature. 
At intermediate chemical potentials, in the region where the chiral phase 
transition would take place if the analysis was restricted to homogeneous 
condensates, we find that two-dimensional modulations are disfavored against 
one-dimensional ones,
indicating that their greater kinetic energy cost is not sufficiently 
compensated by a larger gain in condensation energy.
We also find that a hexagonal structure is even less favored than 
a square lattice in this regime.

From these observations,
it seems unlikely that a phase
where the chiral condensate is modulated along three spatial dimensions
could become thermodynamically favored,
due to its even higher kinetic-energy cost.
Moreover, since a GL analysis has revealed that also close to the Lifshitz 
point phases with higher-dimensional modulations are disfavored against 
one-dimensional ones~\cite{Abuki:2011}, we do not expect that 
higher-dimensional structures appear at finite temperature, at least in 
mean-field approximation.
A numerical analysis to confirm these expectations would of course be 
desirable.

The situation gets successively reversed when we 
increase the chemical potential. At a certain value, we find that
the structure of the ground state changes from having one-dimensional 
modulations to a two-dimensional square lattice and, at even higher $\mu$, 
to an hexagonal shape.
Although the results must not be trusted blindly in this high-density regime, 
which lies at the edge of the expected range of validity of the model, 
we find it nevertheless noteworthy that we reproduce qualitatively 
the behavior recently proposed for quarkyonic 
matter~\cite{Kojo:2011}\footnote{Quasicrystalline structures with discrete 
rotational symmetries higher than six, which have also been discussed in 
Ref.~\cite{Kojo:2011}, are much more difficult to implement in our model
since we heavily rely on the translational periodicity to perform our 
numerical calculations. These structures are anyway expected to appear only 
at even higher chemical potentials.}. 
It is also interesting that a similar sequence of phases was predicted
for a 2D superconductor in a magnetic field \cite{Matsuda:2007}.
On the other hand, the chemical potentials where the two-dimensional 
structures appear in our model belong to the realm of color 
superconductivity. 
Therefore it would be important to include the effects of 
diquark pairing as well.

\section*{Acknowledgments}

We thank Y.~Hidaka, T.~Kojo, L.~McLerran, D.~Nickel and J.~Wambach for 
illuminating discussions. 
This work was partially supported by
the Helmholtz Alliance EMMI, 
the Helmholtz International Center for FAIR, 
and by the Helmholtz Research School for Quark Matter Studies H-QM.

\end{document}